# MULTIDIMENSIONAL WEB PAGE EVALUATION MODEL USING SEGMENTATION AND ANNOTATIONS

K.S.Kuppusamy[1] and G.Aghila[2]

[1]Department of Computer Science, School of Engineering and Technology, Pondicherry University, Pondicherry, India

kskuppu@gmail.com

[2]Department of Computer Science, School of Engineering and Technology, Pondicherry University, Pondicherry, India

aghilaa@yahoo.com

## ABSTRACT

*The evaluation of web pages against a query is the pivot around which the Information Retrieval domain revolves around. The context sensitive, semantic evaluation of web pages is a non-trivial problem which needs to be addressed immediately. This research work proposes a model to evaluate the web pages by cumulating the segment scores which are computed by multidimensional evaluation methodology. The model proposed is hybrid since it utilizes both the structural semantics and content semantics in the evaluation process. The score of the web page is computed in a bottom-up process by evaluating individual segment's score through a multi-dimensional approach. The model incorporates an approach for segment level annotation. The proposed model is prototyped for evaluation; experiments conducted on the prototype confirm the model's efficiency in semantic evaluation of pages.*

## 1. INTRODUCTION

The Web search engines have become the corner-stone of World Wide Web by facilitating efficient access to resources. The information explosion problem has made the users struggle in locating the required information from a huge collection. The resources of the World Wide Web are exponentially growing with each passing day [1]. The resources of the World Wide Web are undergoing the modification and deletion process at a very fast rate [2].

With these exponentially growing resources, the required information identification process has become more complex. Hence it becomes mandatory to efficiently evaluate the web pages against the user supplied query, incorporating the user's informational requirement context. The usefulness of resources of World Wide Web differs across the users. The profile of the user would serve as an important tool in computing the relevance of a resource for the user's current informational requirement context.

This paper proposes an approach towards evaluation of web pages against a user supplied query, by incorporating various dimensions in the evaluation process. The web pages differ from normal text documents due to their inherent ability of providing a structure and linkage to other resources. The proposed model harnesses these structural semantics. A variation of the model termed as MUSEUM (Multi Dimensional Segment Evaluation Model ) proposed by us earlier [3] is utilized in evaluation using multiple dimensions.

The proposed model incorporates the annotations identified through the help of Ontology, in evaluation of the query for content semantics. The objectives the proposed model is as listed below:

- Proposing an approach for evaluating web pages using segment evaluation in a hybrid manner.
- Extending the MUSEUM model with content semantics and exploring the benefits gained out of it.

The remainder of this paper is organized as follows: In Section 2, some of the related works carried out in this domain are explored. Section 3 deals with the proposed model's mathematical representation and algorithms. Section 4 is about prototype implementation and experiments. Section 5 focuses on the conclusions and future directions for this research work.

## 2. RELATED WORKS

This section provides a list of various related works carried out in this domain which formed the motivation for this research work. The proposed model incorporates the following two major fields of study:

- Annotations
- Web Page Segmentation

### 2.1 Annotations

The term annotation refers to attaching additional data to an existing resource [4]. The annotations can be added either manually or through systems. The manual annotations can be easily achieved with existing tools [5]. Though manual annotation is simple, it has certain problems like amount of training required and the understanding of the domain by annotator etc [6].

There exist many Ontology based annotation tools [7][8][9][10][11]. The work in [12] explores and evaluates various annotation tools using multiple criteria. The annotations increases the resources chance to get identified correctly in the required informational context. The semantics of the semantic annotation is explored in [13]. The work explored in [14] discusses eight questions regarding the semantic annotation process. For the proposed model the OpenCalais service is utilized in deriving the context annotations which are further evaluated by the model.

### 2.2 Web Page Segmentation

Web page segmentation is an active research topic in the information retrieval domain in which a wide range of experiments are conducted. Web page segmentation is the process of dividing a web page into smaller units based on various criteria. The following are four basic types of web page segmentation methods. They are Fixed length page segmentation, DOM based page segmentation, Vision based page segmentation and Combined / Hybrid method

A comparative study among all these four types of segmentation is illustrated in [15]. Each of above mentioned segmentation methods have been studied in detail in the literature. Fixed length page segmentation is simple and less complex in terms of implementation but the major problem with this approach is that it doesn't consider any semantics of the page while segmenting. In DOM base page segmentation, the HTML tag tree's Document Object Model would be used while segmenting. An arbitrary passages based approach is given in [16]. Vision based page segmentation (VIPS) is in parallel lines with the way, humans views a page. VIPS [17] is a popular segmentation algorithm which segments a page based on various visual features.

Apart from the above mentioned segmentation methods a few novel approaches have been evolved during the last few years. An image processing based segmentation approach is illustrated in [18]. The segmentation process based on text density of the contents is explained in [19]. The graph theory based approach to segmentation is presented in [20]. The proposed model employs a hybrid segmentation model combining DOM and text density.

## 3. THE MODEL

This section explores about the building blocks of the proposed model for semantically evaluating a page score in a hybrid manner involving both structural semantics and content semantics. The block diagram of the proposed model is as shown in Fig.1. The components of the proposed model are as explained below:

- Segmentor: The Segmentor receives the source page as input and segments it in to a collection. The proposed model employs Document Object Model combined with text density based Segmentation .

- Segment Pool: Segment pool component would store the segments which are identified by the Segmentor component.

- Segment Evaluator: The segment evaluator's role is to evaluate the segment weights. The proposed model employs a variation of the MUSEUM (Multi dimensional Segment Evaluation Model). This model evaluates the segments by considering various structural semantics which includes the following:

    o Link Scorer: The link scorer evaluates the hyperlinks present in the segment against the user supplied query.
    o Image Scorer: The Image scorer is responsible for evaluating the images present in the segment by considering the image metadata items in the page.
    o Visual Scorer: The visual scorer performs the tasks of evaluating the contents of the segment by considering the special visual markup with which the contents are provided. The weight for individual visual markup items is fetched from the VMWT (Visual Markup weight Table).
    o Profile Scorer: The profile scorer's role is to evaluate the profile terms with the contents of the segment. The profile of the user shall be maintained with a weighted list.
    o Freshenss Scorer: The freshness scorer would weigh the fresh contents in the segments, if the page's previous snapshots are available.
    o Theme Scorer: The theme scorer would evaluate the segment's relevance by comparing it with the theme of the page which can be identified using the page's title component.

- Semantic Annotation Service: The semantic annotation service's role is to annotate the segment with various contextual parameters. These contextual parameters provide additional information about the contents of the segment. To fetch the context annotations the OpenCalais [21] web-service is used in the proposed model. The model is defined in a such way that the OpenCalais can be replaced with any similar service in future.

- Context Annotations: The context annotations are the output of Semantic Annotation Service.

- Annotation Evaluator: The annotation evaluator would evaluate the annotations for the current segment. The context annotations are evaluated against the query and the user profile terms to weigh the segment's content semantic relevancy score.

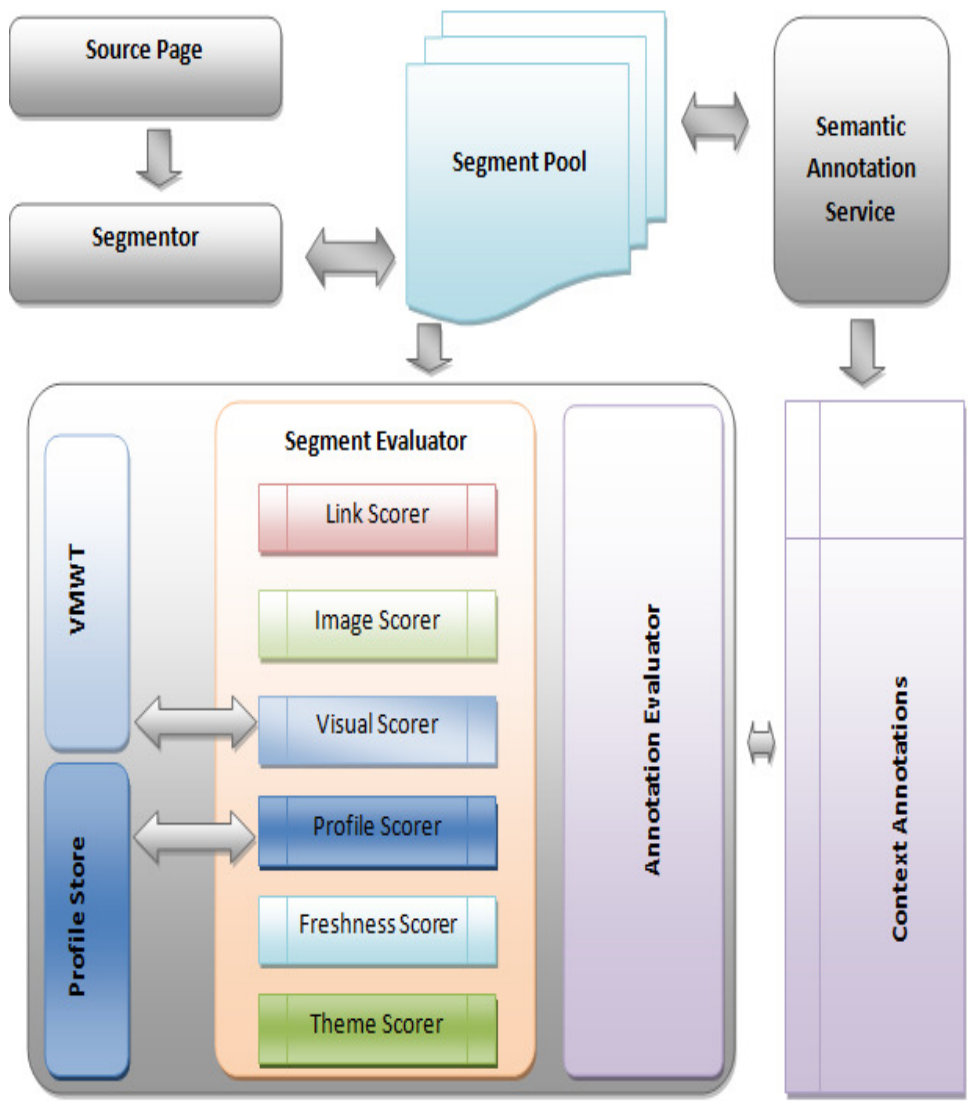

Figure 1.  Block Diagram of proposed Model

### 3.1 Mathematical Model

The proposed model splits the source page into various segments.

$$\Omega = \left\{ \omega_1, \omega_2, \omega_3 ... \omega_n \right\} \tag{1}$$

The user profile terms are stored in the profile store

$$\Gamma = \left\{ \alpha_1, \alpha_2, \alpha_3 ... \alpha_n \right\} \tag{2}$$

In (2), each $\alpha_i$ represents a profile term corresponding to the user. Each of the segments identified in (1), would be evaluated to yield a score. The evaluation is done using a variation of the MUSEUM model. Each segment is evaluated using various structural semantics of the segment.

$$\Phi = \left\{ \forall_{i = 1..n} \delta(\omega_i, \Gamma, Q) \right\} \tag{3}$$

In (3), the function $\delta(\omega, \Gamma, Q)$ indicates the calculation of segment score. The segment score is calculated using five different dimensions as shown in (4).

$$\delta(\omega_i, \Gamma, Q) = \begin{cases} \forall L : \mu(L_i, \Gamma, Q) & \oplus \\ \forall I : \mu(I_i, \Gamma, Q) & \oplus \\ \forall T : \mu(T_i, \Gamma, Q) & \oplus \\ \forall V : \mu(V_i, \Gamma, Q) & \oplus \\ \forall F : \mu(F_i, \Gamma, Q) \\ \forall P : \mu(P_i, \Gamma, Q) & \oplus \end{cases} \tag{4}$$

In (4), $L$ indicates the links, $I$ indicates the Images, $T$ indicates theme of the page, $V$ indicates the visual features, $F$ indicates the freshness weight and $P$ indicates profile weight. The function $\mu$ indicates the score calculation with respect to the corresponding dimension. In (4) $Q$ represent the query terms The symbol $\oplus$ in each row in (4) indicates the summation of scores.

The contextual annotations are fetched from the OpenCalais web service as indicated in (5)

$$\Psi = \left\{ \forall_{i = 1..n} \lambda(\omega_i) \right\} \tag{5}$$

In (5), the function $\lambda$ returns the annotations for the contents of the segment $\omega$. The contextual annotations $\lambda$ consists of various entity categories. The score for each of those categories are calculated as shown in (6)

$$\Psi_S = \left\{ \frac{\forall_{i = 1..n;\ j = 1..m} \lambda_j(\omega_i)}{\forall_{i = 1..n};\ Q_i \otimes \forall_{j = 1..n};\ P_j} \right\} \tag{6}$$

In (6), $\lambda_j(\omega_i)$ represent the context entity "j" for segment "i". The score is calculated by matching it against the terms in the query indicated as $Q_i$ and profile terms indicated as $P_j$. The operator $\otimes$ indicates the fusion of query terms and profile terms.

The final score for the segment is computed by the summation of scores found in (6) and (4) as shown in (7).

$$Score(\omega_i) = \delta(\omega_i, \Gamma) + \Psi_S \qquad (7)$$

By combining the structural semantics and the contextual annotations based content semantics the segment is evaluated in a hybrid manner. The final score of the page is computed by adding the score of the individual segments.

The proposed model utilizes the divide-and-conquer approach of splitting a big problem into smaller ones. The big problem of the page score computation is solved by splitting it in to segment score computation.

## 3.2 The Algorithm

The algorithmic representation of the proposed model is depicted in this section. The proposed model

---

Algorithm PageScore

Input: Source Web Page $\Omega$ , profile $\Gamma$ , Query Q

Output : Score of the page

Begin

   1.  Split the source page into various segment

$\Omega = \{\omega_1, \omega_2, \omega_3 \ldots \omega_n\}$

   2.  Fetch the user profile terms

$\Gamma = \{\alpha_1, \alpha_2, \alpha_3 \ldots \alpha_n\}$

   3.  for each segment $\omega_i$

        call SegmentScore ( $\omega_i$ , $\Gamma$ , Q)

        update page score

   4.  return pagescore

End

---

The algorithm PageScore receives the webpage, profile terms and query as input. The source page is segmented. For each segment of the source page the SegmentScore algorithm computes the segment weight.

Algorithm AnnotationScore

Input: Source Segment $\omega$ , profile $\Gamma$ , Query Q

Output :Score of the segment

Begin

   1.  Call OpenCalais WebService with $\omega$ contents

   2.  For each category retrieved in (1)

        Calculate categoryScore

        Update Score

---

Algorithm SegmentScore

Input: Source Segment $\omega$ , profile $\Gamma$ , Query Q

Output :Score of the segment

Begin

   1.  compute the link weight $\mu(L_i,\Gamma,Q)$

   2.  compute the image weight $\mu(I_i,\Gamma,Q)$

   3.  compute the theme weight $\mu(T_i,\Gamma,Q)$

   4.  compute the visual feature weight $\mu(V_i,\Gamma,Q)$

   5.  compute the freshness weight $\mu(F_i,\Gamma,Q)$

   6.  compute the profile weight $\mu(P_i,\Gamma,Q)$

   7.  call AnnotationScore( $\omega$ , $\Gamma$,$Q$)

   8.  sum up the weight components of (1) to (7)

   9.  return weight

End

The algorithm PageScore receives the webpage, profile terms and query as input. The source page is segmented. For each segment of the source page the SegmentScore algorithm computes the segment weight. The SegmentScore algorithm employs a multi-dimensional approach in computing the structural semantics score.

The contextual annotation score is computed by the AnnotationScore algorithm which employs a web service to fetch the annotations. In the contextual annotations received, each entity category is evaluated for its weight.

## 4. EXPERIMENTS AND RESULT ANALYSIS

This section explores the experimentation and results associated with the proposed model for evaluating web pages using a combination of multi-dimensional structural semantics and contextual annotations. The prototype implementation is done with the software stack including Ubuntu Linux, Apache, MySql and PHP. The OpenCalais Web-service call is made through PHP using cURL library [22]. With respect to the hardware, a Core i5 processor system with 3 GHz of speed, 8 GB of RAM is used. The internet connection used in the experimental setup is a 128 Mbps leased line.

The screenshot in Fig.2 depicts a web page with segment boundaries highlighted. The context annotations retrieved is shown in Fig. 3.

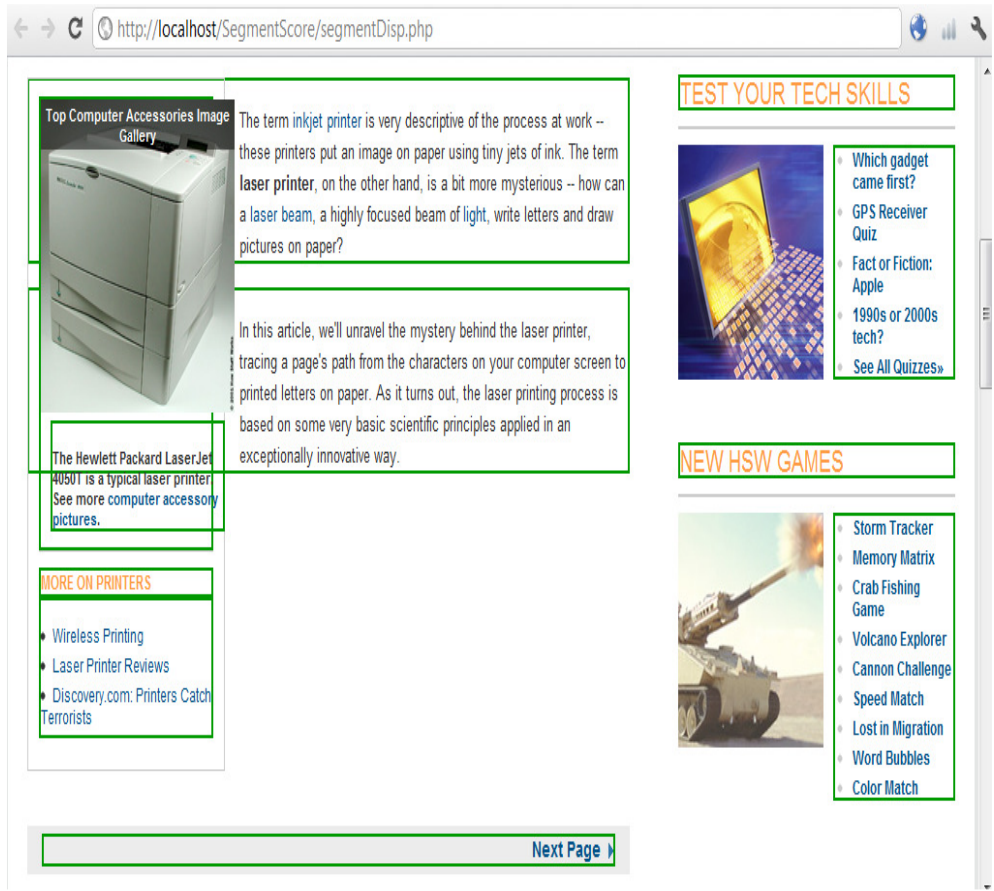

Figure 2. Web Page with Segment Boundaries Highlighted

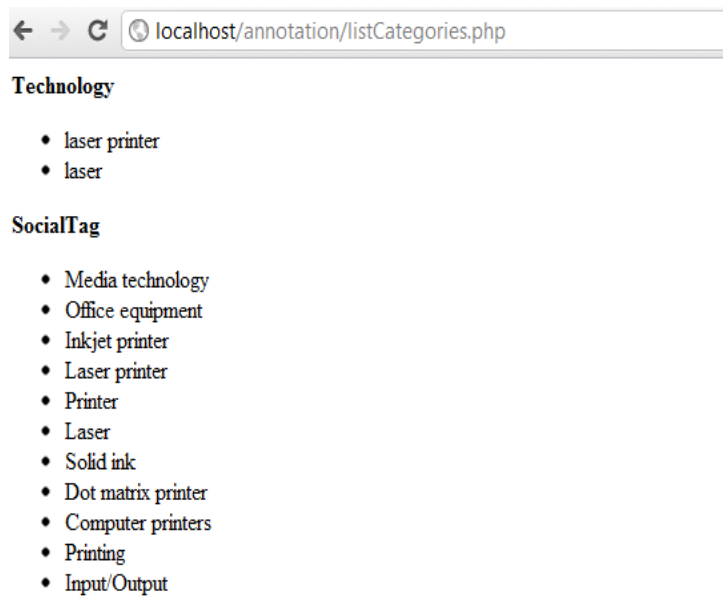

Fig. 3. Context Annotations for a Sample Segment

The experiments were conducted in various sessions. The result data of the experiments are tabulated in Table 1.

Table 1.  Experimental Results

| Session ID | MSC | MSSS | MCAS |
|---|---|---|---|
| 1 | 23.21 | 6.32 | 4.74 |
| 2 | 17.16 | 7.25 | 5.44 |
| 3 | 16.45 | 8.35 | 6.26 |
| 4 | 12.77 | 9.34 | 7.01 |
| 5 | 18.31 | 12.51 | 9.38 |
| 6 | 12.45 | 10.26 | 7.7 |
| 7 | 22.35 | 11.33 | 8.5 |
| 8 | 23.12 | 12.55 | 9.41 |
| 9 | 17.54 | 13.12 | 9.84 |
| 10 | 20.32 | 15.13 | 11.35 |
| 11 | 18.35 | 13.21 | 9.91 |
| 12 | 19.65 | 14.32 | 10.74 |
| 13 | 18.35 | 15.51 | 11.63 |
| 14 | 17.11 | 14.34 | 10.76 |
| 15 | 13.45 | 14.65 | 10.99 |

In Table 1 MSC stands for mean segment count which indicates the mean of the number of segments in that session. MSSS stands for Mean Structural Semantic Score and MCAS stands for Mean Contextual Annotation Score. In Fig 2, the chart depicts the comparison of various segment parameters computed during the page score evaluation process.

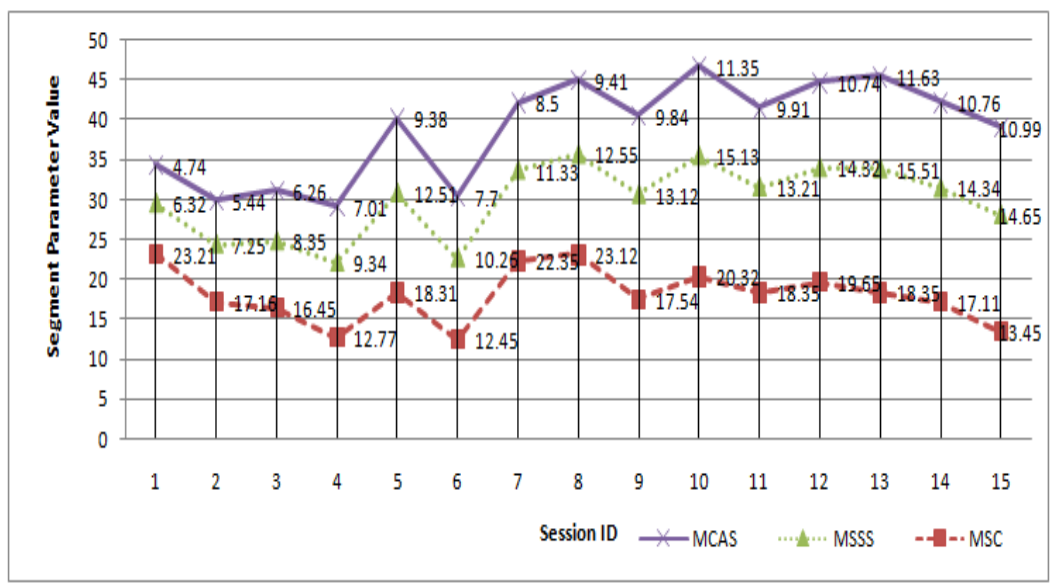

Figure 4. Comparison of Segment Parameters across Sessions

The experiments conducted on the prototype provide few interesting trends. It was observed that the structural semantic score computation and the context annotation based score were directly proportional to each other. In session where the MSSS was less, the value of MCAS was also low. Across the sessions the value of MSSS was 11.87 and MCAS was 8.91. The experiments suggested that adding the semantic annotation increased the segment score proportionately to leading to efficient evaluation of segments at the micro-level and the complete web page at the macro-level.

## 5. CONCLUSIONS AND FUTURE DIRECTIONS

This section lists out the conclusions derived from the proposed model for hybrid evaluation of web pages with semantic segment annotations and multi-dimensional structural semantics evaluation. The following is the list of conclusions:

- The webpage evaluation shall be enriched by performing a hybrid evaluation with both structural semantics and content semantics. In the proposed research work, the score of the segments improved by 57.14% across the sessions.

- The user profile terms which got combined with search query terms produced better evaluation scores.

The future directions for the proposed model is as listed below:

- The proposed model shall be further enriched to handle other languages apart from English in the evaluation process.

- The evaluation process shall be enhanced by incorporating domain specific requirements in the evaluation process.

**Authors**

K.S.Kuppusamy is an Assistant Professor at Department of Computer Science, School of Engineering and Technology, Pondicherry University, Pondicherry, India. He has obtained his Masters degree in Computer Science and Information Technology from Madurai Kamaraj University. He is currently pursuing his Ph.D in the field of Intelligent Information Management. His research interest includes Web Search Engines, Semantic Web. He has made more than ten peer reviewed international publications. 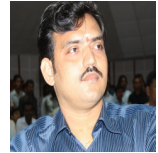

G. Aghila is a Professor at Department of Computer Science, School of Engineering and Technology, Pondicherry University, Pondicherry, India. She has got a total of 22 years of teaching experience. She has received her M.E (Computer Science and Engineering) and Ph.D. from Anna University, Chennai, India. She has published more than 55 research papers in web crawlers, ontology based information retrieval. She is currently a supervisor guiding 8 Ph.D. scholars. She was in receipt of Schrneiger award. She is an expert in ontology development. Her area of interest includes Intelligent Information Management, artificial intelligence, text mining and semantic web technologies. 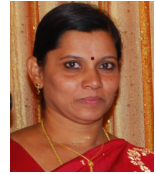